\documentclass[aps,prb,twocolumn,floatfix,amsmath,amssymb,superscriptaddress]{revtex4}
\usepackage[final]{graphicx}
\usepackage{bm}
\usepackage{afterpage}
\usepackage{color}
\usepackage{comment}
\usepackage[colorlinks=true,urlcolor=blue,linkcolor=blue,citecolor=blue]{hyperref}
\usepackage{amsmath,amssymb,bbold,bm}

\begin{document}

\title{Quantum Monte Carlo study of the attractive kagome-lattice Hubbard model}

\author{Xingchuan Zhu}
\affiliation{Interdisciplinary Center for Fundamental and Frontier Sciences, Nanjing University of Science and Technology, Jiangyin, Jiangsu 214443, P. R. China}

\author{Wanpeng Han}
\affiliation{School of Physics, Beihang University,
Beijing, 100191, China}

\author{Shiping Feng}
\affiliation{ Department of Physics,  Beijing Normal University, Beijing, 100875, China}

\author{Huaiming Guo}
\email{hmguo@buaa.edu.cn}
\affiliation{School of Physics, Beihang University,
Beijing, 100191, China}

\begin{abstract}
Recent experimental discovery of several families of kagome-lattice materials has boosted the interest in electronic correlations on kagome
lattice. As an initial step to understand the observed complex phenomena, it is helpful to know the correspondence between simple forms of interactions and the induced correlated states on kagome lattice. Considering the lack of such studies, here we systematically investigate the attractive kagome-lattice Hubbard model using the mean-field approach and determinant quantum Monte Carlo (DQMC). A charge-density-wave order satisfying the triangle rule is predicted by the mean-field treatment, and subsequent DQMC simulations provide indirect evidence for its existence. The $s$-wave superconductivity is found to be stabilized at low temperatures, and exists in dome regions of the phase diagrams. We then determine the superconducting critical temperature quantitatively by finite-size scaling of the pair structure factor. These results may be helpful in understanding the observed superconductivity in kagome-lattice materials.
\end{abstract}


\maketitle
\section{Introduction}
The kagome lattice, formed by corner sharing triangles, is unique in that it combines the intriguing physics of geometry frustration,
flat band and Dirac fermions, and thus sets an ideal platform for novel quantum phases~\cite{mekata2003kagome}. Due to the strong geometry frustration, the antiferromagnetic spin-1/2 Heisenberg model on the kagome lattice is a paradigmatic realization of a quantum spin liquid (QSL). Many efforts have been devoted to uncover its physical nature among a gapless $U(1)$ Dirac, a gapped topological $Z_2$, and chiral QSLs\cite{balents2010spin,yan2011spin,Savary_2016,RevModPhys.89.025003,doi:10.1126/science.aay0668,RevModPhys.88.041002}.
The rich features in the energy dispersion of the itinerant electrons on the kagome lattice have arose great interest in investigating the exotic quantum orders of fermions. Especially, the spin-orbit coupling can open a nontrivial gap at the Dirac point and the quadratic band crossing point touching the flat band, generating $Z_2$ topological insulating states\cite{guohm2009}. Remarkably, several magnetically ordered materials that contain a kagome lattice have been found recently, and the experimental evidences point to the realization of the above simple topological model\cite{ye2018massive,liu2018giant,yin2018giant,PhysRevLett.121.096401,kang2020dirac}. Theoretically, a rich variety of interaction-driven phases have been proposed, including: dynamically-generated topological phase, various spin or charge bond orders and density waves\cite{PhysRevB.81.235115,PhysRevB.82.075125,PhysRevB.90.035118,PhysRevLett.110.126405}, and the superconducting instability\cite{PhysRevB.85.144402,PhysRevB.86.121105,PhysRevB.87.115135,PhysRevB.94.014508}.

The interest in electronic correlations in kagome lattice is further boosted by the recent experimental discovery of several families of kagome materials,
 such as: $\mathrm{T}_m\mathrm{X}_n$ ($\mathrm{T= Fe, Co}$ and $\mathrm{X= Sn, Ge}$)
 and $\mathrm{A}\mathrm{V}_3\mathrm{Sb}_5$ ($\mathrm{A= Cs, K, Rb}$)\cite{ye2018massive,liu2018giant,ortiz2019}.
 The exhibited topological quantum states and a cascade of correlated phases have received significant research interests\cite{jiang2021kagome,doi:10.1063/5.0079593,neupert2022charge}.
 Specifically, in the new kagome prototype materials $\mathrm{A}\mathrm{V}_3 \mathrm{Sb}_5$ ($\mathrm{A= Cs, K, Rb}$),
 stacked ideal kagome network of vanadium layers give rise to rich correlated electronic phases including:
 charge density wave $(\mathrm{CDW})$ order occurring below $T_c^{\mathrm{CDW}} \approx$ $80-110 \mathrm{~K}$,
 a further transition at $T'=35K$ with an additional unidirectional charge ordering vector,
 and unconventional superconductivity with critical temperature $T_c \approx 0.9-2.7 \mathrm{~K}$\cite{zhao2021cascade}.
 The CDW order, which may be closely related to van Hove singularities at the Fermi level\cite{kang2022twofold},
 exhibits exotic characteristics such as: time-reversal symmetry breaking\cite{jiang2021unconventional,FENG20211384,mielke2022time}
 and nematicity\cite{nie2022charge}. Its interplay with superconductivity has been investigated by applying external pressure.
 As the CDW is destabilized by the pressure, the superconducting state undergoes an unconventional two-dome evolution in the critical temperature,
 which suggests a complex intertwinement of the CDW state and superconductivity\cite{PhysRevLett.126.247001,yu2021unusual}.
 At present, the microscopic interacting mechanism underlying the above correlated states is challenging, and remains elusive\cite{PhysRevLett.127.177001,PhysRevLett.127.046401,PhysRevB.103.L241117}.
 First, it is helpful to know the correspondence between simple forms of interactions and their induced symmetry-breaking orders on kagome lattice. Nevertheless, till now, the prototype models of interacting fermions on kagome lattice are still less investigated than their counterparts on square and honeycomb geometries\cite{meng2010quantum,sorella2012absence,PhysRevX.3.031010,PhysRevX.6.011029,PhysRevB.72.085123,PhysRevB.91.165108,PhysRevB.105.075118,PhysRevB.105.245131,PhysRevB.104.L121118,PhysRevB.104.165127}.

In this paper, we perform a systematic study of attractive kagome-lattice Hubbard model,
with the aim of estimating the relevance of the attractive on-site interaction to the experimental discoveries.
We first analyze the physical property of possible CDW orders at $\rho=2/3$, and perform a mean-field (MF) study of the CDW phase transition.
Then DQMC is applied to unveil the correlated phases therein. From the charge correlation function, it is found the instability to CDW patterns satisfying the triangle rule may occur at the
Dirac points. Next we calculate the $s$-wave pair structure factor, and map out the phase diagrams in the $(\mu,T)$ plane. Although the finite-size effect is
apparent in small lattices, robust superconducting (SC) domes exits for large values of $U,L$.
Finally, we determine the SC transition temperature using finite-size scaling. These results suggest that the $s$-wave superconductivity supported
by some experiments in $\mathrm{A}\mathrm{V}_3\mathrm{Sb}_5$ may originate from electronic attractive on-site interaction.

This paper is organized as follows. Section II introduces
the model we will investigate, along with our computational
methodology. Section III presents the MF theory for the CDW transition. Section IV uses DQMC simulations to study
the possible CDW state at $2/3$ filling and superconductivity with on-site pairing. Section VI gives the conclusions.

\section{The model and method}

\begin{figure}[htbp]
\centering \includegraphics[width=8.5cm]{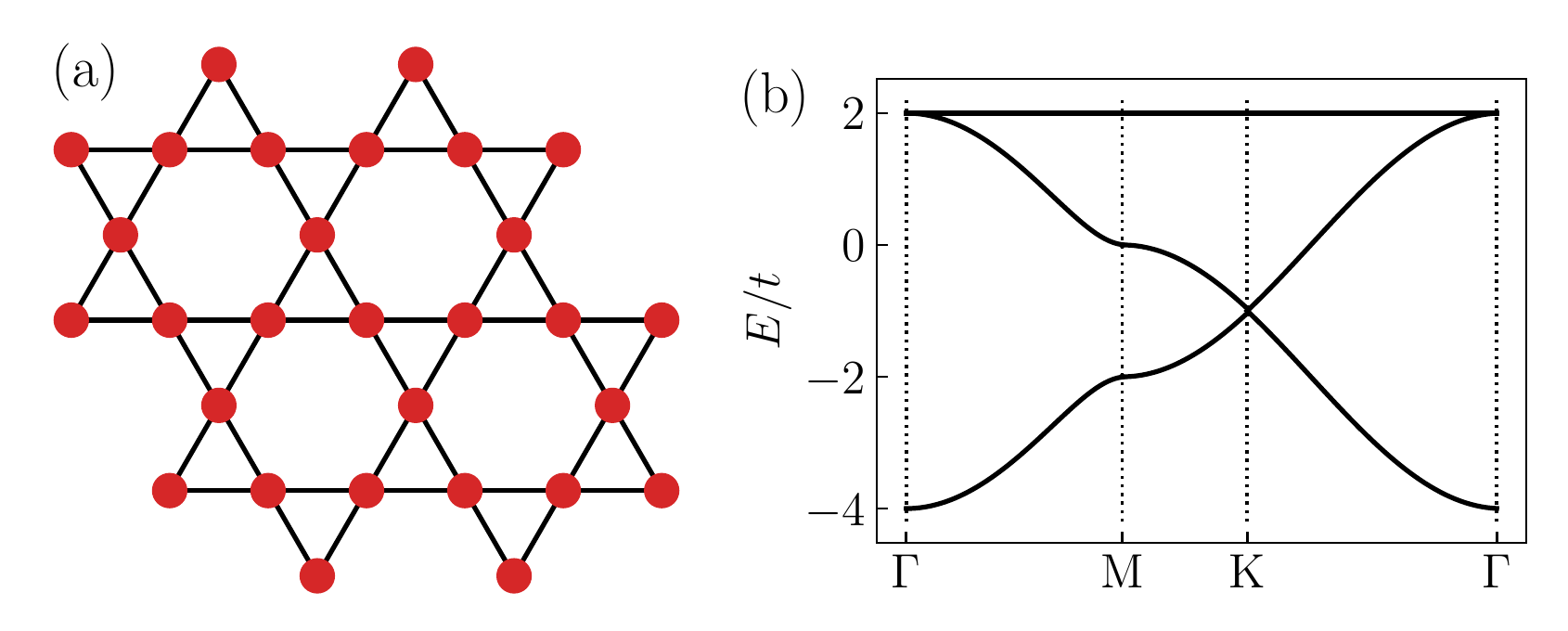} \caption{(a) The geometry of the kagome lattice, which is a triangular Bravais
lattice with a three-site unit cell. (b) The band structure along the high-symmetry directions in the
Brillouin zone.}
\label{fig1}
\end{figure}

We start from the attractive kagome-lattice Hubbard model,
\begin{align}
H=-t\sum_{\langle i j\rangle \sigma} c_{i \sigma}^{\dagger} c_{j \sigma}^{\phantom{\dagger}}-U \sum_{i}\left(n_{i \uparrow}-\frac{1}{2}\right)\left(n_{i \downarrow}-\frac{1}{2}\right),
\label{eq:H}
\end{align}
where $c_{i \sigma}^{\dagger}$ and $c_{i \sigma}$ are the creation and annihilation operators, respectively, at site $i$ with spin $\sigma=\uparrow, \downarrow$; $\langle ij\rangle$ denotes nearest neighbors; $n_{i \sigma}=c_{i \sigma}^{\dagger} c_{i \sigma}$ is the number of electrons of spin $\sigma$ on site $i$, and $U$ is the on-site attractive interaction. Throughout the paper, the hopping amplitude is set to $t = 1$ as the unit of energy.

The kagome lattice has a three-site unit cell as shown in Fig.~\ref{fig1}(a). In momentum space, the Hamiltonian at $U=0$ is given by~\cite{guohm2009}
\begin{align}
\mathcal{H}_{0}({\mathbf{k}})=-2 t\left(\begin{array}{ccc}
0 & \cos k_{1} & \cos k_{3} \\
\cos k_{1} & 0 & \cos k_{2} \\
\cos k_{3} & \cos k_{2} & 0
\end{array}\right),
\end{align}
where $k_n={\mathbf{k}}\cdot {\mathbf{a}}_n$ (the sublattice index $n=1,2,3$) with ${\mathbf{a}}_1=(1,0),{\mathbf{a}}_2=(-1,\sqrt{3})/2$, and ${\mathbf{a}}_3=-({\mathbf{a}}_1+{\mathbf{a}}_2)$. The spectrum of $\mathcal{H}_{0}({\mathbf{k}})$ has one flat band $E_{3}({\mathbf{k}})=2t$ and two dispersive ones,
\begin{align}
E_{1,2}({\mathbf{k}})=t[-1\pm \sqrt{4f({\mathbf{k}})-3}],
\end{align}
with $f({\mathbf{k}})=\cos^2 k_1+\cos^2 k_2+\cos^2 k_3$. Bands $1$ and $2$ touch at two inequivalent Dirac points ${\bf K}_{\pm}=(\pm 2\pi/3,0)$ at energy $-t$ shown in Fig.~\ref{fig1}(b). For $\frac{1}{3}$ filling, the lowest band is filled, and the low-energy excitations resemble those of graphene, which are linear, $\epsilon_{1,2}=\pm \sqrt{3}t|\vec{q}|$, with $\vec{q}=(q_x,q_y)$ a small displacement away from the Dirac points.

At finite interactions, Eq.\eqref{eq:H} is solved numerically via DQMC, where one decouples the on-site interaction term through the introduction of an auxiliary Hubbard-Stratonovich field, which is integrated out stochastically. The only errors are those associated with the statistical sampling, the finite spatial lattice size, and the inverse temperature discretization. These errors are well controlled in the sense that they can be systematically reduced as needed, and further eliminated by appropriate extrapolations. Unlike the repulsive Hubbard model on kagome lattice where the infamous sign problem exists at all densities~\cite{PhysRevB.41.9301,PhysRevLett.94.170201,PhysRevB.92.045110}, the attractive case under our investigation is free of the sign problem\cite{PhysRevLett.66.946,PhysRevLett.62.1407,PhysRevB.48.3976,PhysRevB.80.245118,PhysRevB.69.184501}. This allows DQMC to reach the low temperatures needed to study the ground-state properties. In the following, we use the inverse temperature discretization $\Delta\tau=1/16$, and the lattice has $N=3\times L \times L$ sites with $L$ up to $12$.


\section{The MF theory}

To explore possible CDW orders at $\rho=2/3$, we first investigate the physical properties of the CDW order preserving the translation symmetry of the kagome lattice. The following CDW term is added to the non-interacting Hamiltonian in Eq.(2),
\begin{align}
\mathcal{H}_{CDW}({\mathbf{k}})=\textrm{diag}(w_1,w_2,w_3),
\end{align}
where $w_{l}(l=1,2,3)$ represents the on-site potential of the $l$-th sublattice. Since it is independent of spin, we can discard the spin index, and focus on the spinless case at $\rho=1/3$.

Here a central concern is whether the above CDW can open up a gap at the Dirac points. This is more easily revealed based on the low-energy Hamiltonian, which can be obtained by linearizing ${\cal H}_{\bf k}={\cal H}_{0}({\bf k})+{\cal H}_{CDW}({\bf k})$ near ${\bf K}_{\pm}$ and subsequently projecting onto the subspace associated with the lowest two bands. With the above procedure, we find the following low-energy Hamiltonian,
\begin{align}
h_{\ell}(\mathbf{k})=v\left[\sigma_z\left(k_x-\mathcal{A}_x^{\ell}\right)+\sigma_x\left(k_y-\mathcal{A}_y^{\ell}\right)\right]+ {\mathbb{1}} w
\end{align}
for valley $\ell$, with the Fermi velocity $v=\sqrt{3}t$, $w=(w_1+w_2+w_3)/3$, and
\begin{align*}
\mathcal{A}_x^{\ell} &=\left(2 w_2-w_1-w_3\right) \ell / 6 v, \\
\mathcal{A}_y^{\ell} &=\left(w_1-w_3\right) \ell / 6 t.
\end{align*}

Thus CDW couples to the Dirac fermions as a gauge field, which moves the positions of the Dirac point in the Brillouin zone, and does not open up a gap. Nevertheless, when CDW is large enough to make the two Dirac points meet and merge with each other, the system becomes gapped. This is in great contrast to the situation in graphene, where an on-site staggered potential always opens up a gap at the Dirac points.

Then it is helpful to perform a MF analysis of the Hamiltonian Eq.(2) to reveal the possible CDW orders. In the MF approximation, the interaction term in Eq(2) can be decoupled as,
\begin{align}\label{decoupled}
n_{i,\uparrow} n_{i,\downarrow}= \langle n_{i,\uparrow} \rangle n_{i,\downarrow} + n_{i,\uparrow} \langle n_{i,\downarrow} \rangle - \langle n_{i,\uparrow} \rangle \langle n_{i,\downarrow} \rangle.
\end{align}
There may be various kinds of CDW phases at the low filling $\rho=2/3$. Here we consider a set of well-established CDW orders in the literature satisfying the triangle rule: each unit cell of the kagome lattice only contains one electron-rich site, and it is always surrounded by electron-poor sites at nearest neighbors. There are still multiple such CDW orders, and herein we focus on a simple configuration in which one specific sublattice is occupied by majority electrons.
To incorporate the above CDW order, the average of the number operator writes as $\langle n_{i,\sigma} \rangle= \rho_{_{i}}$ with $\rho_{_{i}}$ being the order parameter. Since the assumed CDW preserves the translation symmetry of the kagome lattice, $\rho_i$ may only differ within the unit cell, and takes three values $\rho_{_l} (l=1,2,3)$. Then the attractive Hubbard interaction becomes,
\begin{align}\label{HubbardtermMF}
- U \sum_i n_{i,\uparrow} n_{i,\downarrow} =  - U \sum_{l=1,2,3} \sum_{i \in l} \rho_{_l} n_i+E_0,
\end{align}
where $n_i=n_{i,\uparrow} + n_{i,\downarrow}$ is the operator of total number of electrons,  and $E_0 = \frac{NU}{3} (\rho_1^2 + \rho_2^2 + \rho_3^2)$ with the total number of sites $N$ that is a constant. In the momentum space, the MF Hamiltonian writes as,
\begin{align}\label{HubbardtermMFm}
{\cal H}_{MF}^{\sigma}({\bf k}) = {\cal H}_{0}^{\sigma}({\bf k}) + {\cal H}_{diag},
\end{align}
with
\begin{align}\label{HubbardtermMFmD}
{\cal H}_{diag} =
\begin{pmatrix}
-U \rho_1 & 0 & 0 \\
0 & -U \rho_2 & 0 \\
0 & 0 & -U\rho_3
\end{pmatrix}.
\end{align}
The energy spectrum is directly obtained by diagonalizing the above Hamiltonian. Supposing the low energy band is $E_k$ (degenerate for both spin copies), the total ground-state energy is $E_{tol}=2\sum_{\bf k} E_{\bf k} +E_0$. Minimizing $E_{tol}$ with respect to $\rho_{_{l}}$, we can obtaind the self-consistent equation for the order parameters
\begin{align}
\rho_{_{l}} = -\frac{3}{UN} \frac{\partial \left( \sum_k E_k\right)}{\partial \rho_{_l}}.
\end{align}

\begin{figure}[htbp]
\centering \includegraphics[width=7.5cm]{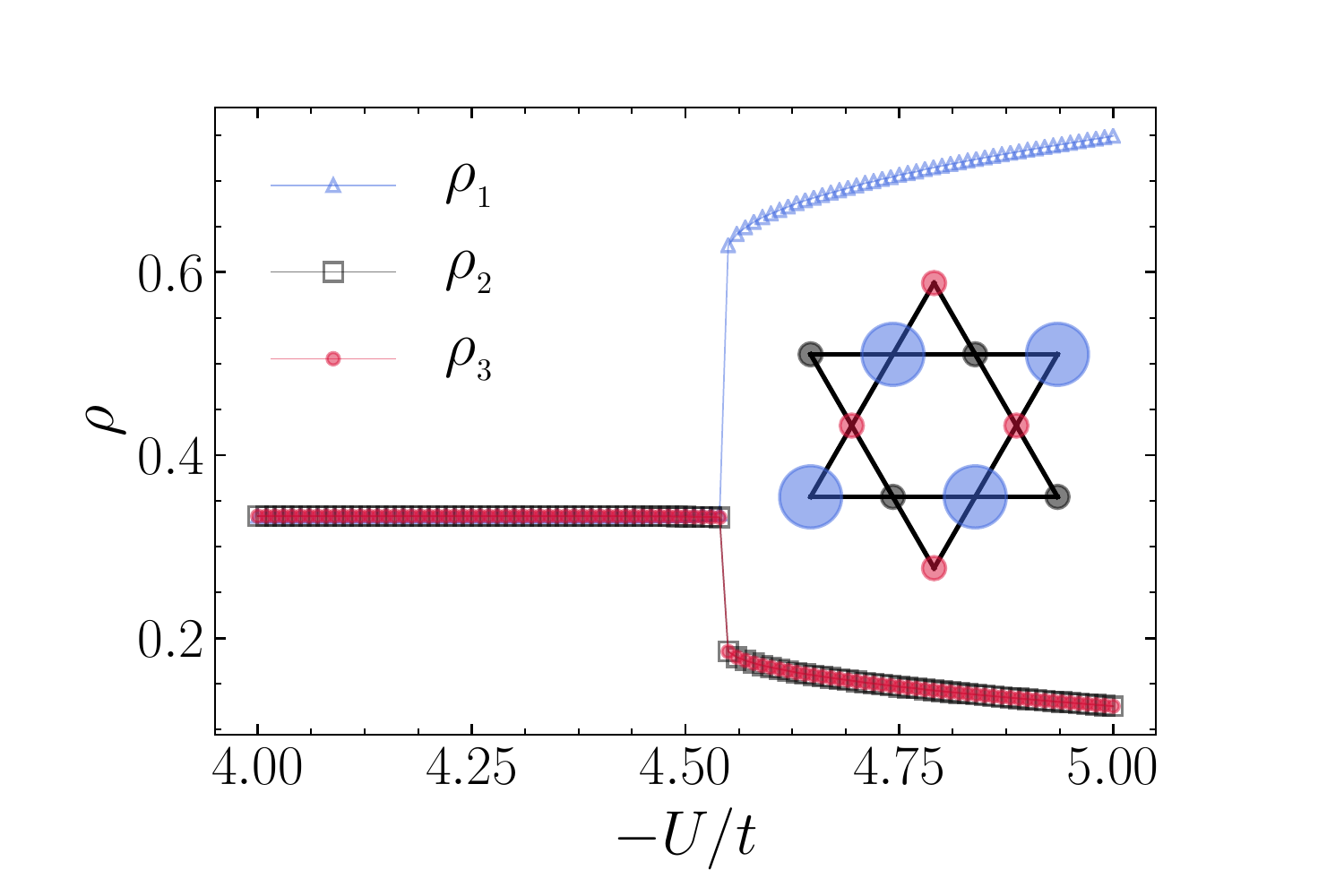} \caption{The mean-field order parameters $\rho_1$, $\rho_2$ and $\rho_3$ as a function of $U$. The curves exhibit a discontinuity marking the CDW phase transition, and the critical interaction is determined to be $U_c/t=-4.54$.}
\label{mf4sites}
\end{figure}

Figure \ref{mf4sites} plots the order parameters $\rho_{_l}$ calculated self-consistently as a function of $U$. The order parameter is uniform, and all equal to $1/3$ at small interactions. Then at a critical strength $U_c/t=4.54$, the curves suddenly split into two branches, and the value of $\rho_1$ becomes much larger than that of $\rho_2, \rho_3 (\rho_2=\rho_3)$, suggesting the occurrence of a CDW phase transition.

\section{The DQMC results}

\subsection{CDW at $2/3$ filling}

\begin{figure}[htbp]
\centering \includegraphics[width=9.cm]{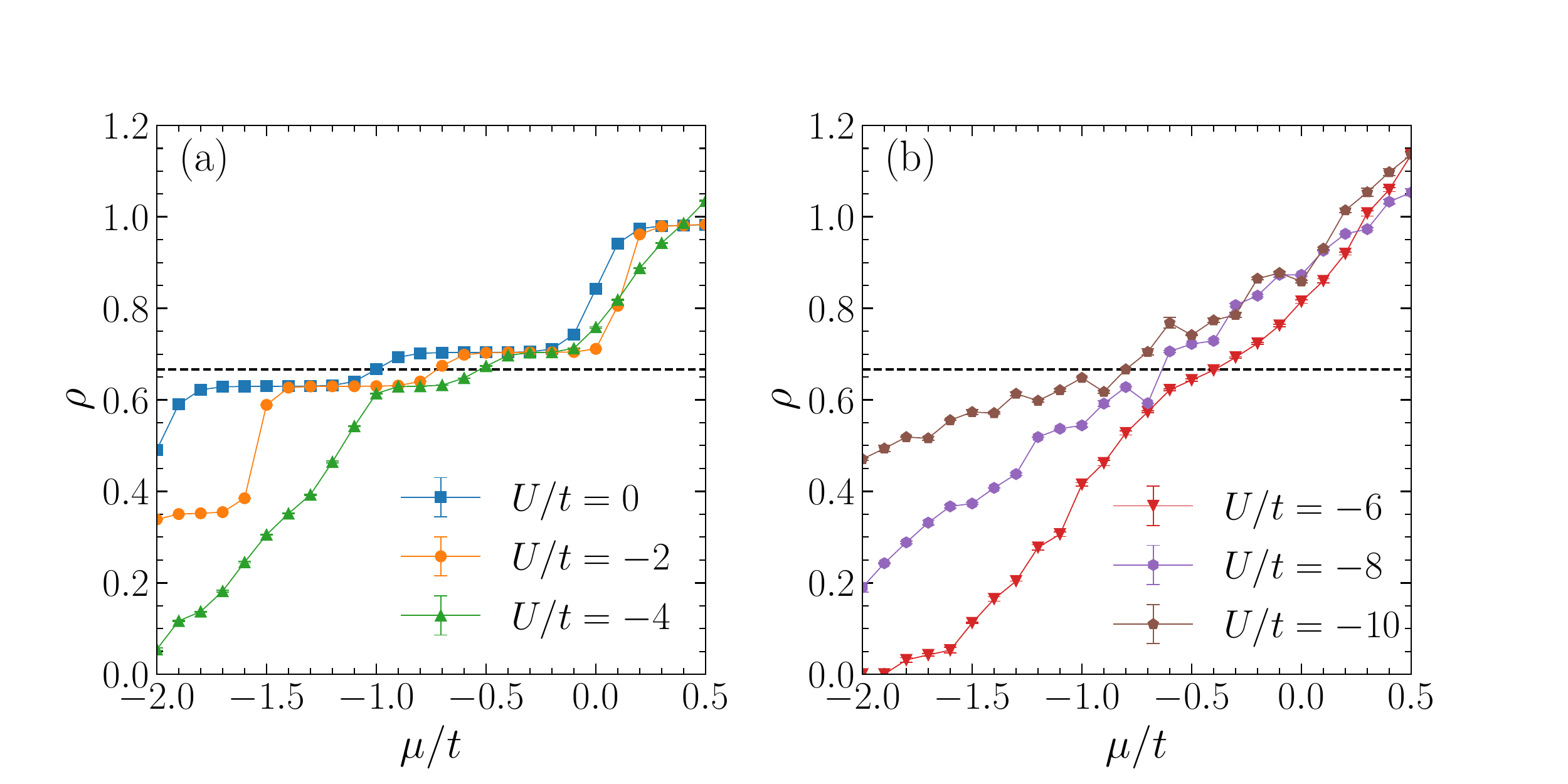} \caption{The average density as a functions of $\mu$ at the attractive interaction (a) $U/t=0,-2,-4$ and (b) $U/t=-6,-8,-10$. Here the lattice size is $L=6$, and the inverse temperature is $\beta t=18$.}
\label{densitybutongU}
\end{figure}

Next we apply DQMC to unveil the physical properties of the Hamiltonian (1) quantitatively. Figure \ref{densitybutongU} plots the average density, $\rho=\frac{1}{N}\sum_{i\sigma}\langle n_{i\sigma}\rangle$, versus $\mu$ for various values of $U$. There exists evident finite-size plateaus near the Dirac density $\rho=2/3$, which  persists up to $U/t\sim -4$. Afterwards, $\rho$ continuously increases with $\mu$, and the curves show no special features. As discussed above, since the CDW order may not gap out the Dirac points, it is unclear here whether a $\rho=2/3$ CDW has been induced by large attractive interactions.

\begin{figure}[htbp]
\centering \includegraphics[width=9.cm]{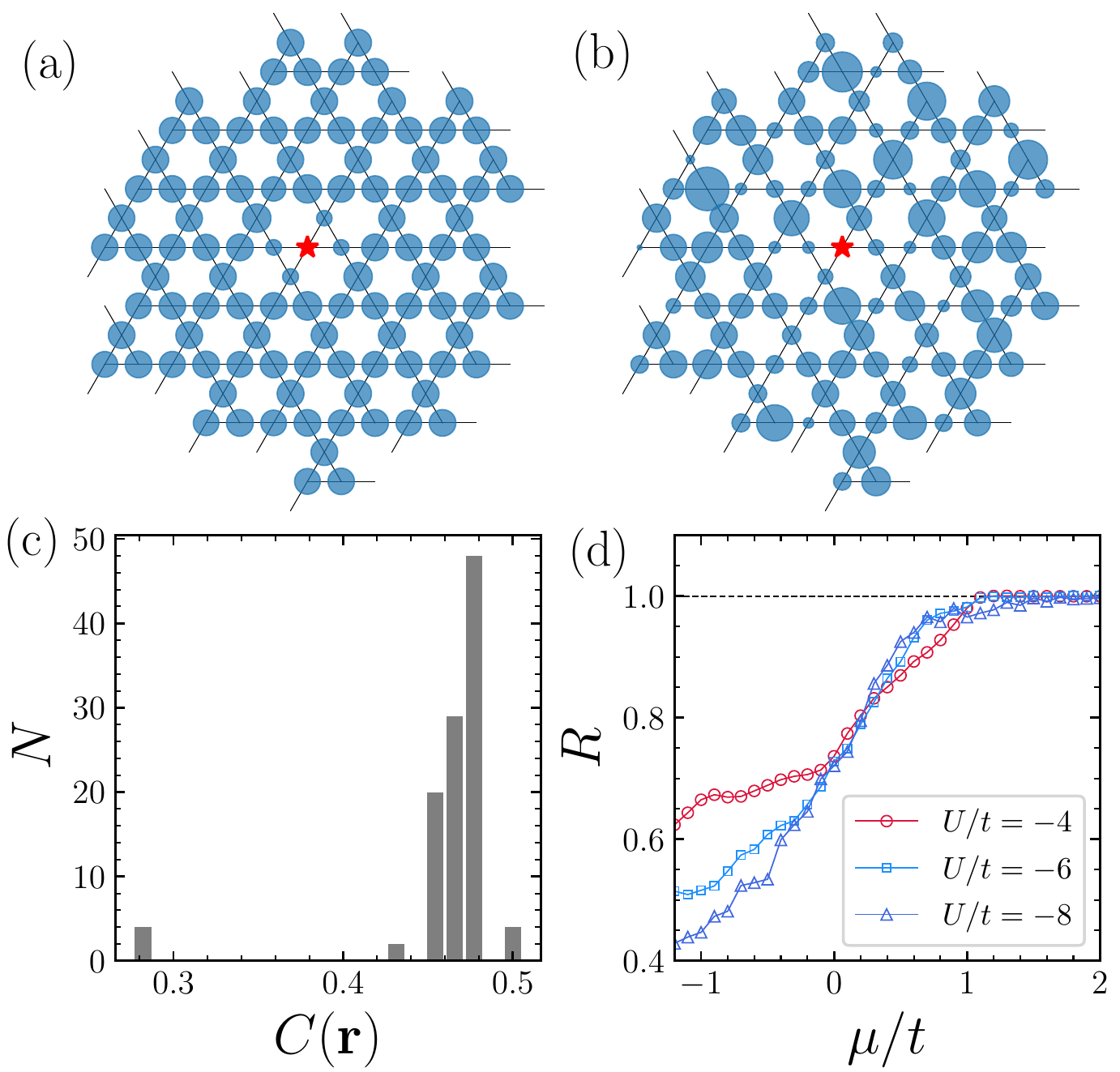} \caption{(a) The charge correlation function $C({\bf r})$ for $\mu/t=-0.4$ (corresponding to $\rho=2/3$) on a $L=6$ lattice. The red star marks the reference site, and the magnitude of the correlation is represented by the radii of the solid blue circle. (b) $C({\bf r})$ in one updated configuration of the DQMC measurement of (a). (c) The distribution of the values of $C({\bf r})$ in (a). (d) The ratio $R=C({\bf r}_{nn})/C({\bf r}_{max})$ as a function of chemical potential. Here the parameters are $U/t=-8$, and $\beta t=12$.}
\label{fig4}
\end{figure}

In order to detect the possible CDW phase, we plot in Fig.~\ref{fig4} the real-space charge-charge correlation function, which is defined as $C({\bf r})=\langle n_{i}n_{i+{\bf r}}\rangle$ with $n_{i}=n_{i\uparrow}+n_{i\downarrow}$. $C({\bf r})$ is nearly uniform over the whole lattice except the nearest-neighbor (NN) charge correlations, whose values are apparently smaller than the other ones. We further calculate the ratio $R=C({\bf r}_{nn})/C({\bf r}_{max})$ with ${\bf r}_{nn} ({\bf r}_{max})$ the NN (maximum) distance in the lattice. It is found that as the average density increases and goes away from $\rho=2/3$, $R$ increases continuously, and becomes uniform from $\mu/t \sim 1$. Although no CDW pattern is identified at $\rho=2/3$, each configuration in the histogram of DQMC measurements has clearly inhomogeneous charge correlations. The above behavior may be due to the multi-fold degeneracy of the CDW phase fulfilling the triangle rule. After averaged over the different charge patterns, the charge correlations becomes uniform. Nevertheless, since the triangle rule always restricts the occupation of the NN sites in all degenerate configurations, the value of the NN charge correlations remains greatly reduced. Hence, our results provides indirect evidence for the existence of CDW patters satisfying the triangle rule.

\subsection{Superconductivity with on-site pairing}


\begin{figure}[htbp]
\centering \includegraphics[width=9.cm]{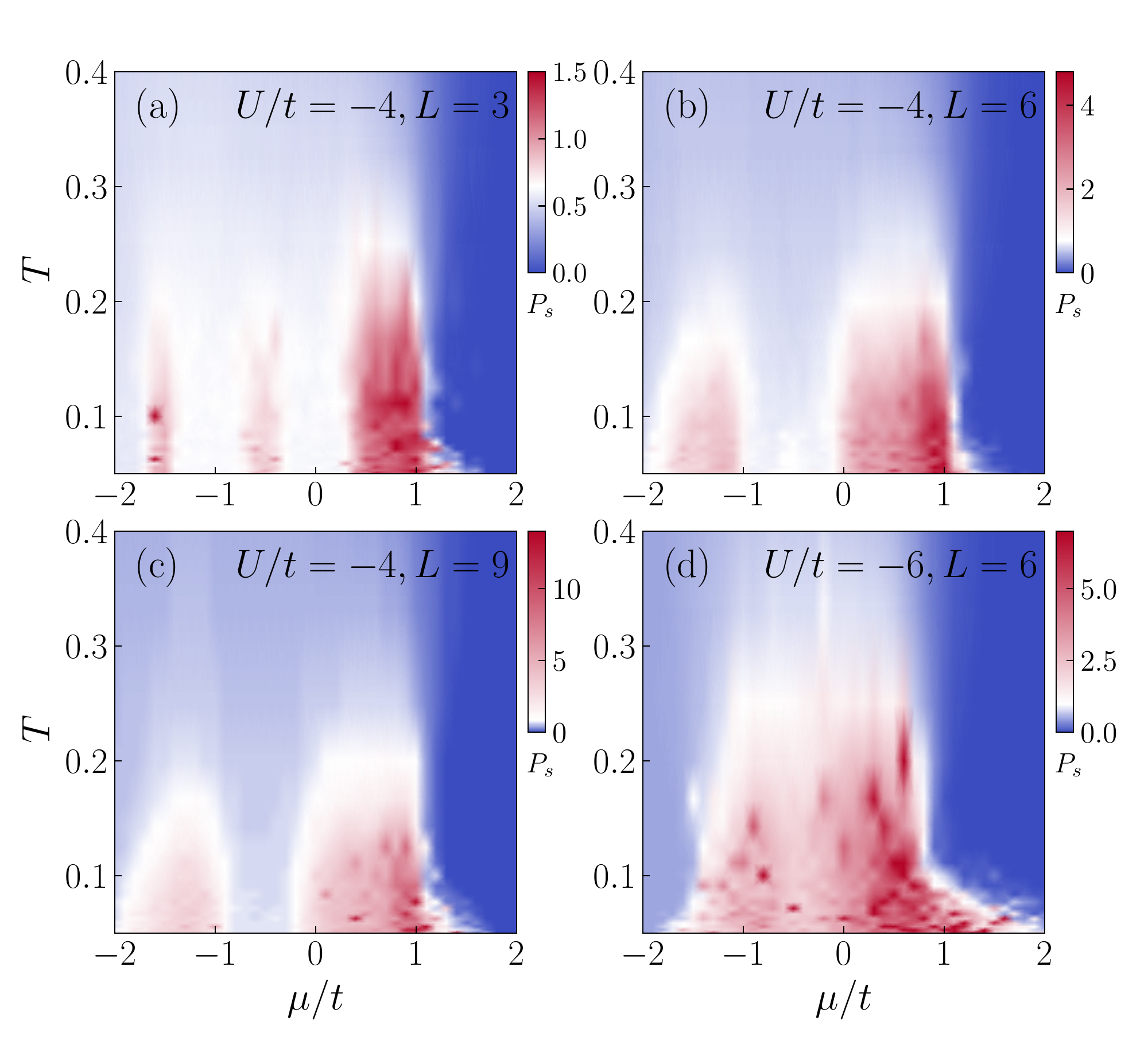} \caption{The pair structure factor $P_s$ in the space of parameters $T/t$ vs. $\mu/t$ at $U/t=-4$ for lattice sizes: (a) $L=3$, (b) $L=6$, and (c) $L=9$. (d) Similar plot with $U/t=-6$ and $L=6$.}
\label{fig5}
\end{figure}

The $s$-wave superconductivity is characterized by the pair structure factor,
\begin{align}
P_s=\left\langle\Delta^{\dagger} \Delta+\Delta \Delta^{\dagger}\right\rangle,
\end{align}
with
\begin{align}
\Delta^{\dagger}=\frac{1}{\sqrt{N}} \sum_i c_{i \uparrow}^{\dagger} c_{i \downarrow}^{\dagger}.
\end{align}
Figure \ref{fig5} plots the pair structure factor in the $(\mu,T)$ plane for several lattice sizes. For small lattice, the finite-size effect is very apparent, which is similar to that has been observed in the attractive Hubbard model on square lattice. As shown in Fig.~\ref{fig5}(a), the $s$-wave pairing is enhanced at sufficiently low temperatures near several special values of $\mu/t$. This behavior has been attributed to the coarse discretization of the Brillouin zone in small lattices, which persists even at moderate interaction $U/t=-4$. As the lattice size increases, there remain two disconnected SC domes. As shown in Fig.~\ref{fig5}(b) and (c), the gap between them decreases with the lattice size, and seems highly related to the flat region of the average density near $\rho=2/3$. For stronger interaction strength $U/t=-6$, there is only one large SC dome in the phase diagram. Correspondingly, there are no visible plateaus in the curve of the average density.

\begin{figure}[htbp]
\centering \includegraphics[width=9.cm]{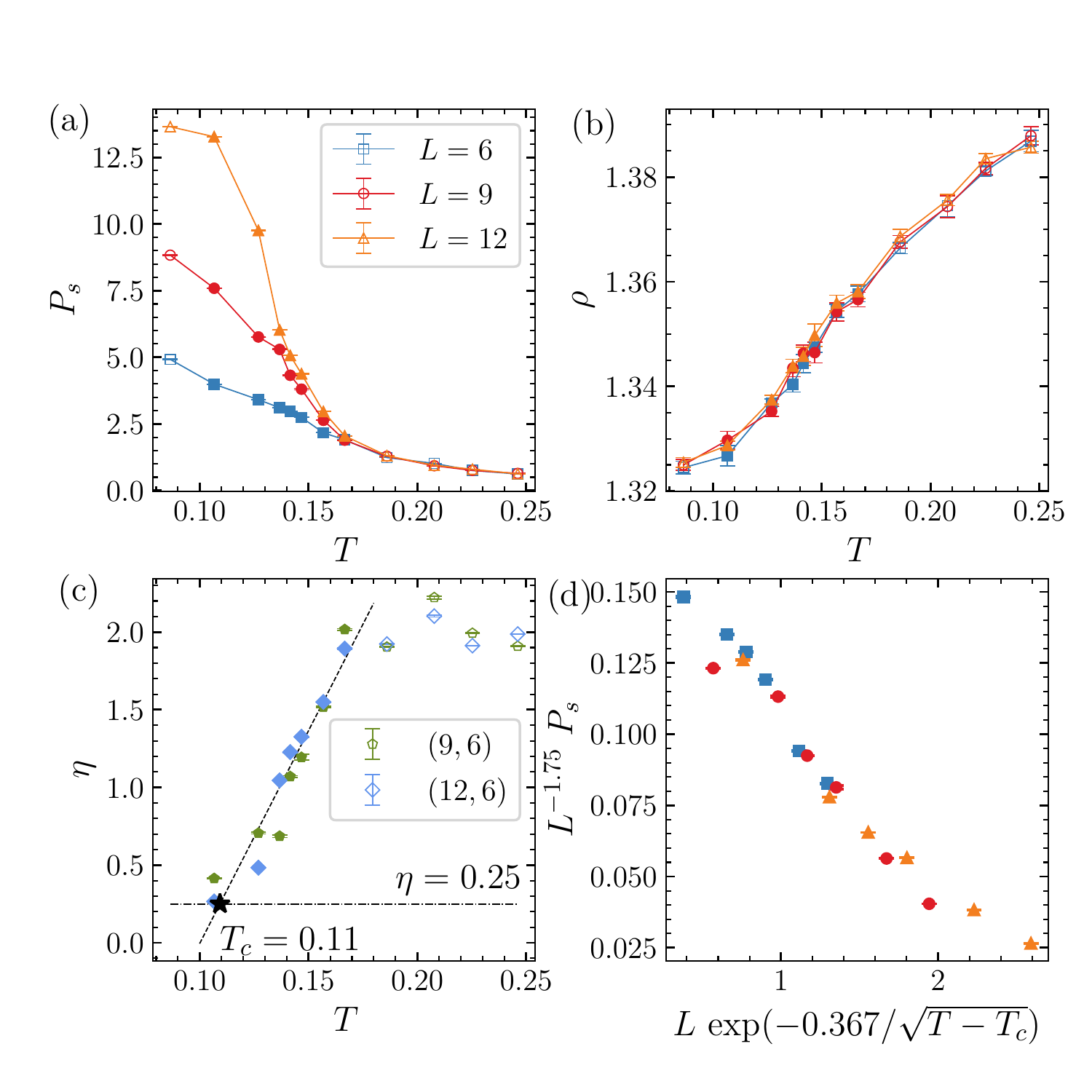} \caption{(a) $P_s$ as a function of temperature $T$ for various lattice sizes. (b) The average density vs. $T$, which correspond to the plots in (a). (c) The exponent $\eta(T)$ extracted according to Eq.(15) as a function of temperature. The critical temperature is determined to be $T_c=0.11$ by the condition $\eta(T_c)=0.25$. (d) The collapse of the curves in (a) using the scaling form Eq.(14) and $T_c$ determined in (c). Here $A=-0.367$ is used to obtain the best collapse. The interaction strength $U/t=-4$ and the chemical potential $\mu/t=0.9$ are used.}
\label{fig6}
\end{figure}

Next we perform a quantitative analysis of the SC critical temperature at $\mu/t=0.9$, where the superconductivity is the most predominant for $U/t=-4$. As the temperature is lowered, the pair structure factor increases monotonically [see Fig.~\ref{fig6}(a)]. For high temperatures, $P_s$ is size-independent due to the absence of SC long-range order. Conversely, $P_s$ increases significantly with the lattice size at low temperatures, which is a hallmark of the occurrence of SC state. The usual way to investigate the properties of the Hubbard model is to fix the average density. However, since DQMC works in a grand-canonical ensemble, the above routine has an increased overhead to determine the chemical potential that produces the desired filling. Here we choose to fix the chemical potential for different temperatures. In the temperature range of interest ($T/t=0.1-0.16$), this routine results in a slight deviation of the densities around $\rho=1.35$ [see Fig.~\ref{fig6}(b)].

One expects the decay of the real-space correlations follows
\begin{align}
C(r) \equiv\left\langle c_{\mathbf{i} \uparrow}^{\dagger} c_{\mathbf{i} \downarrow}^{\dagger} c_{\mathbf{j} \downarrow} c_{\mathbf{j} \uparrow}+\text { H.c. }\right\rangle \sim r^{-\eta(T)},
\end{align}
where $r=|\mathbf{i}-\mathbf{j}|$. Then the pair structure factor scales as
\begin{align}
P_s=L^{2-\eta\left(T\right)} f(L / \xi),
\end{align}
with the coherence length $\xi\sim \textrm{exp}[-A/(T-T_c)^{\frac{1}{2}}]$~\cite{PhysRevLett.66.946,PhysRevLett.62.1407,PhysRevB.69.184501,mondaini2022universality}.
Here $\eta(T)$ is temperature-dependent, and can be extracted by dividing the above scaling form from two different lattice sizes $L, L'$. The obtained exponent writes as
\begin{align}
\eta(T)=2-\frac{\ln \left[P_s(L, T) / P_s\left(L^{\prime}, T\right)\right]}{\ln \left(L / L^{\prime}\right)}.
\end{align}
We take a $L=6$ lattice as the reference one, and the extracted $\eta(T)$ at each temperature according to the above equation is illustrated in Fig.~\ref{fig6}(c). At high temperatures, the pair structure factor has negligible size-dependence, thus the exponent $\eta(T)$ saturates around $2$ in this regime. Otherwise, in the $T\rightarrow T_c$ limit, $\eta(T_c)=0.25$ is expected. Above $T_c$, $\eta(T)$ increases monotonically to the saturated value $2$. By a linear fit of the increasing regime, the critical temperature is determined to be $T_c/t=0.11$ for $U/t=-4$ and $\mu/t=0.9$. Subsequently, we collapse $P_s$ of different lattice sizes using the scaling form in Eq.(14) with the above $T_c$ and $A$ being adjusted to give the best data collapse. As shown in Fig.~\ref{fig6}(d), the collapse onto a single curve is pretty good for the $\eta(T)$-increasing region.

\section{Conclusions}
We investigate the attractive kagome-lattice Hubbard model with two complementary methods: the MF theory and large-scale DQMC simulations. The MF analysis predicts a CDW transition, with the configuration of the CDW order satisfying the triangle rule. Subsequent DQMC simulations provides indirect evidence for its existence at strong interactions. Then, by calculating the pair structure factor, the s-wave superconductivity is shown to be stabilized at low temperatures, and exists in dome regions of the phase diagrams. We finally determine the SC critical temperature quantitatively by finite-size scaling of the pair structure factor.

The pairing symmetry is important to understand the SC mechanism in $\mathrm{A}\mathrm{V}_3\mathrm{Sb}_5$. Its two aspects, i.e., gap structure and the nature of electron pairing state, have been much investigated experimentally. Unexpectedly, various techniques have yielded inconsistent results, including: singlet or triplet electron pairing, nodeless or node gap function\cite{Mu_2021,ni2021anisotropic,PhysRevLett.127.187004,duan2021nodeless,zhao2021nodal,PhysRevX.11.031026}. The complexity may be due to the multi-band nature of the SC state, and it is still challenging to reconcile the apparently contradictory observations. Nevertheless, our results suggest a $s$-wave pairing mechanism by the attractive Hubbard interaction, which may be helpful in understanding the complex SC phenomena in kagome-lattice materials.

\section*{Acknowledgments}
The authors thank Fan Yang and Wen Yang for helpful discussions. H.G. acknowledge support from the National Natural Science Foundation of China (NSFC) grant Nos.~11774019 and 12074022, the NSAF grant in NSFC with grant No. U1930402. S.F.~is supported by the National Key Research and Development Program of China, and NSFC under Grant Nos.~11974051 and 12274036. X.Z. is supported by the Fundamental Research Funds for the Central Universities (Grant No. AE89991/383).


\appendix


\renewcommand{\thefigure}{A\arabic{figure}}

\setcounter{figure}{0}

\bibliography{ddirac}

\end{document}